\documentclass{nature}

 {\usepackage[utf8]{inputenc}}           % switch to utf8

\usepackage[USenglish]{babel}
\usepackage{comment}
\usepackage{dcolumn}   % needed for some tables
\usepackage{bm}        % for math
\usepackage{amssymb}   % for math
\usepackage{amsmath}
\usepackage{verbatim} 
\usepackage{color}
\usepackage[section] {placeins}
\usepackage{graphicx}  % needed for figures
\usepackage[caption=false,font=footnotesize]{subfig}
\usepackage[left]{lineno}

\title{Terahertz-driven linear electron acceleration }

\author{Emilio A. Nanni$^{1}$, Wenqian R. Huang$^{1}$, Kyung-Han Hong$^{1}$, Koustuban Ravi$^{1}$, Arya Fallahi$^{2,3}$, Gustavo Moriena$^4$, R. J. Dwayne Miller$^{3,4,5}$ \& Franz X. K{\"a}rtner$^{1,2,3,6}$}

\begin{document}

\maketitle

\begin{affiliations}
 \item Department of Electrical Engineering and Computer Science, Research Laboratory of Electronics, Massachusetts Institute of Technology, Cambridge, MA 02139, USA
 \item Center for Free-Electron Laser Science and The Hamburg Center for Ultrafast Imaging, Hamburg, Germany
 \item Deutsches Elektronen Synchrotron, Ultrafast Optics and X-rays Division, Hamburg, Germany
 \item Department of Chemistry and Physics, University of Toronto, Toronto, Canada
 \item Max Planck Institute for the Structure and Dynamics of Matter, Hamburg, Germany
\item Department of Physics, University of Hamburg, Hamburg, Germany
\end{affiliations}
%\begin{linenumbers}
\begin{abstract}

The cost, size and availability of electron accelerators is dominated by the achievable accelerating gradient. Conventional high-brightness radio-frequency (RF) accelerating structures operate with 30-50 MeV/m gradients\cite{emma2010first,ishikawa2012compact}. Electron accelerators driven with optical or infrared sources have demonstrated accelerating gradients orders of magnitude above that achievable with conventional RF structures\cite{geddes2004high,hafz2008stable,plettner2008proposed,kneip2009near,peralta2013demonstration,varin2002acceleration,payeur2012generation}. However, laser-driven wakefield accelerators require intense femtosecond sources and direct laser-driven accelerators and suffer from low bunch charge, sub-micron tolerances and sub-femtosecond timing requirements due to the short wavelength of operation. Here, we demonstrate the first linear acceleration of electrons with keV energy gain using optically-generated  terahertz (THz) pulses. {THz-driven accelerating structures enable high-gradient electron or proton accelerators with simple accelerating structures, high repetition rates and significant charge per bunch\cite{yoder2005side,hebling2011optical,palfalvi2014evanescent}}. Increasing the operational frequency of accelerators into the THz band allows for greatly increased accelerating gradients due to reduced complications with respect to breakdown and pulsed heating. Electric fields in the GV/m range have been achieved in the THz frequency band using all optical methods\cite{vicario2014gv}. With recent advances in the generation of THz pulses via optical rectification of slightly sub-picosecond pulses, in particular improvements in conversion efficiency\cite{huang2013high} and multi-cycle pulses\cite{chen2011generation}, increasing accelerating gradients by two orders of magnitude over conventional linear accelerators (LINACs) has become a possibility. These ultra-compact THz accelerators with extremely short electron bunches hold great potential to have a transformative impact for free electron lasers, future linear particle colliders, ultra-fast electron diffraction, x-ray science, and medical therapy with x-rays and electron beams. 

\end{abstract}

At RF frequencies where conventional sources (klystrons, \textit{etc.}) are efficient, surface electric field gradients in accelerating structures are limited by RF induced plasma breakdown. Empirically, the breakdown threshold has been found\cite{kilpatrick1957criterion,wang1989rf} to scale as $E_s \propto {f^{1/2}}/{\tau^{1/4}}$ where ${{E}_{s}}$ is the surface electric field, $f$ is the frequency of operation and $\tau$ is the pulse length indicating that higher frequencies and shorter pulse durations are clearly beneficial. {This empirical scaling is limited by the onset of field-emission due to the electric field on material surfaces. For most common accelerating materials this onset is in the 10's of GV/m range\cite{forbes2007,thompson2008breakdown}.} Additionally, low frequencies (\textit{i.e.} GHz) inherently require long RF pulses because a single RF cycle is long (on the order of ns) and traditional sources work most efficiently when operating over a very narrow frequency spectrum (\textit{i.e.} long pulse length). Practically, this results in a significant amount of average power coupled into the structure if a high repetition rate is used. 

If high accelerating gradients are desired, presently available ultrafast near-IR terawatt and petawatt laser technology  based on chirped pulse amplification allows for multi-GeV/m gradients\cite{peralta2013demonstration, breuer2013laser,carbajo2014ultra}.  {However,  due to the short wavelength, infrared optical pulses prove difficult to use for the direct acceleration of electrons with significant charge per bunch, which is an important parameter. For example, the coherent emission in an FEL from an electron bunch scales with the square of the number of electrons in the bunch\cite{emma2010first}.} In order to prevent emittance growth and increased energy spread, the electron bunch needs to occupy a small fraction of the optical cycle. Even for a long-infrared wavelength of 10~$\mu$m, 1$^o$ of phase in the optical wavelength corresponds to only $\sim$28 nm. Another practical concern would be the timing precision between the optical cycle and the arrival of the electron bunch. For example, 1$^o$ phase jitter, commonly required for operational accelerators, requires less than {100 attosecond timing jitter between the optical pulse and the electron bunch, which is challenging to maintain over extended distances.} Difficulties increase further when considering the available options for guiding the optical light in order to decrease the phase velocity to match the electron beam. A guided mode at a wavelength of 10~$\mu$m would require sub-micron precision for aligning the electron bunch and the optical waveguide. 

Alternatively, laser-plasma wakefield accelerators have demonstrated GeV/m accelerating gradients \cite{faure2004laser, geddes2004high, hafz2008stable} with 100 TW - PW low repetition rate sources with percent-level energy spread and jitter for the electron bunch. The complexity of these plasma-wakefield accelerators is significant because the acceleration mechanism relies on plasma bubbles which are easily subject to instabilities. Also, power scaling of the required high-energy femtosecond pulse sources is challenging.

THz frequencies provide the best of both worlds. On one hand the wavelength is long enough that we can fabricate waveguides with conventional machining techniques, provide accurate timing and accommodate a significant amount of charge per bunch. At 0.3~THz the wavelength is 1 mm and 1$^o$ of phase precision corresponds to 10~fs timing jitter, which is readily achievable\cite{schibli2003attosecond}. On the other hand, {the frequency is high enough that the plasma breakdown threshold for surface electric fields increases into the multi-GV/m range\cite{thompson2008breakdown}.} Additionally, using optical generation techniques we can have very short THz pulses ($\leq$100~ps) generated by picosecond lasers readily available at high average power and under rapid development. These short pulses allow for a limited amount of pulsed heating and a limited amount of average power loading at high repetition rates (on the order of kHz and above). Both the increase in operational frequency and reduction in pulse length will play a role in increasing the breakdown limit. 

Here we report the first experimental demonstration of linear electron acceleration using an optically-generated 10~$\mu$J single-cycle THz pulse centered at 0.45~THz {(See Methods: THz Pulse Generation)}. The THz pulse accelerates electrons in a circular waveguide consisting of a dielectric capillary with a metal outer boundary. The dielectric slows the group and phase velocity of the THz wave allowing it to accelerate low energy electrons. A schematic view of the THz accelerator is shown in Figure~1(a) with a photograph of the THz LINAC in Figure~1(b). Using 60~keV electrons, from a DC electron gun, an energy gain of 7~keV  is observed in a 3~mm interaction length. The single-cycle THz pulse, see Figure 1(c-e), is produced via optical rectification of a 1.2~mJ, 1.03~$\mu$m laser pulse with a 1~kHz repetition rate {(See Methods: Pump Source)}. The THz pulse, whose polarization is converted from linear to radial by a segmented waveplate {(See Methods: Radially Polarized THz Beam)}, is coupled into a waveguide with 10~MV/m peak on-axis electric field {(See Methods: Structure Testing)}. A 25~fC input electron bunch is produced with a 60~keV DC photo-emitting cathode excited by a 350~fs UV pulse {(See Methods: UV Photoemitter)}. {The accelerating gradient in the THz structures demonstrated in this work can be as high as GeV/m with a single cycle THz pulse of 10~mJ (See Methods: THz LINAC), which can be readily produced by a 250~mJ IR pulse when using optimized THz generation\cite{huang2013high}. Laser systems producing such and even higher energy ps pulses with up to kHz repetition rates are on the horizon\cite{zapata2014cryogenic}.} 

%Previously reported work numerically showed that a dielectric-loaded waveguide was a good candidate for an acceleration structure\cite{wong2013compact}.
In this experiment, the THz waveguide supports a traveling TM$_{01}$ mode that is phase-matched to the velocity of the electron bunch produced by the DC photoinjector. {It is the axial component of the TM$_{01}$ mode which accelerates the electrons as they co-propagate down the waveguide.} A traveling-wave mode is advantageous when considering the available single-cycle THz pulse because it does not require resonant excitation of the structure. A dielectric-loaded circular waveguide was selected due to the ease of fabrication in the THz band\cite{mitrofanov2011reducing}.  The inner diameter of the copper waveguide is 940~$\mu$m with a dielectric wall thickness of 270~$\mu$m. This results in a vacuum space with a diameter of 400~$\mu$m. The significant thickness of the dielectric is due to the low energy of the electrons entering the structure, and will decrease significantly at higher energy. One critical aspect for THz electron acceleration is proper interaction between the electron beam and the THz pulse.  Coupling the radially polarized THz pulse into the single mode dielectric waveguide was achieved with a centrally loaded dielectric horn. The design was optimized to maximize coupling with minimal fabrication complexity. Finite element electromagnetic simulations with \textsc{HFSS} %\cite{HFSS}
 indicate excellent coupling of the THz pulse over a $\sim$200~GHz bandwidth, which is compatible with the bandwidth of the radially-polarized mode converter. The accelerating waveguide is 10~mm in length, including a single tapered horn for coupling the THz into the waveguide. Alignment between the THz waveguide and the DC gun is provided by a pin-hole aperture in a metal plate with a diameter of 100~$\mu$m that abuts the waveguide. The THz pulse is coupled into the waveguide downstream of the accelerator and it propagates along the full length of the waveguide before being reflected by the pin-hole aperture, which acts as a short at THz frequencies. After being reflected the THz pulse co-propagates with the electron bunch. {The low initial energy of the electrons results in the rapid onset of a phase-velocity mismatch between the electron bunch and the THz pulse once the electrons have been accelerated by the THz pulse and this limits the interaction length to 3~mm (See Methods: THz LINAC).} 

The electron beam energy is determined via energy-dependent magnetic steering with a dipole located after the accelerator. Figure~2(a-b) shows images of the electron beam produced by the micro-channel plate detector. The measured energy spectrum from the electron bunch with and without THz is shown in Figure~2(c-d) for an initial mean energy of 59~keV. The curves are compared with \textsc{PARMELA} simulation results used to model the DC gun and the THz LINAC. The electron bunch {RMS} length after the pin-hole, $\sigma_z=45~\mu$m, is long with respect to the wavelength of the THz pulse in the waveguide, $\lambda_g=315~\mu$m, resulting in both the acceleration and deceleration of particles. With the available THz pulse energy, a peak energy gain of 7~keV was observed by optimizing the electron beam voltage and timing of the THz pulse. The modeled curve in Figure~2(d) concurred with experiments for an on-axis electric field of 8.5~MV/m. Using this estimated field strength, at the exit of the LINAC, the modeled transverse and longitudinal emittance are 240~nm-rad and 370~nm-rad, respectively. An increase in emittance from a transverse emittance of 25~nm-rad and a longitudinal emittance of 5.5~nm-rad after a pin-hole located at the waveguide entrance is due to the long electron bunch length compared to the THz wavelength and can be easily remedied with a shorter UV pulse length.

The energy gain achieved during the interaction of the electron bunch with the THz pulse is dependent on the initial energy of the electrons, because the setup is operated in the non-relativistic limit where the velocity of the electrons varies rapidly. {If the initial energy is decreased, the particle velocity decreases and the phase-velocity mismatch with the THz pulse increases reducing the interaction length and the acceleration of the particle.} In Figure~3(a) the achieved mean output energy of the electron bunch is shown vs. the initial energy. The peak energy gain is observed at the highest initial energy.  {A single particle model\cite{wong2013compact}} demonstrates good agreement with an on-axis gradient of 2.5~MeV/m. In Figure~3(b) the mean energy of accelerated electrons is shown as a function of the THz pulse delay for 55~keV initial energy. The large temporal range of observed acceleration results from waveguide dispersion which broadens the single cycle pulse temporally as it propagates. Accelerated electrons are observed over the full range of the phase-matched THz pulse due to the long length of the electron bunch as it enters the THz waveguide. Modeling indicates that at the entrance of the THz waveguide for this initial energy the electron bunch full width is 1.5~ps in length which is already a significant fraction of the THz cycle (2.2~ps).
 
In conclusion, optically-generated THz pulses were used to accelerate electrons in a simple and practical THz accelerator for the first time. An energy gain of 7~keV was achieved over a 3~mm interaction length with good modeled emittance. Performance of these structures improves with an increase in electron energy and gradient making them attractive for compact accelerator applications. {With upgrades to pump laser energy and technological improvements to THz sources, laboratory demonstration of GeV/m gradients in THz LINACs is realistic. Multi-GeV/m gradients and $>10$~MeV energy gain are achievable in dielectric-loaded circular waveguides\cite{wong2013compact} with 10~mJ THz pulses and the injection of electrons at relativistic energies (See Methods: THz LINAC). } {The available THz pulse energy scales with IR pump energy, with recently reported results of mJ THz pulse energies and $\sim$3$\%$ IR-to-THz conversion efficiencies\cite{vicario2014generation,fulop2014efficient}.} Multiple stages of THz acceleration can be used to achieve higher energy gain with additional IR pump lasers for subsequent stages. {Timing jitter will improve upon the jitter of conventional accelerators since the accelerating field and photo-emitting pulse are produced by the same drive laser. Therefore, one expects the resulting electron bunch to have tighter synchronization than possible in today’s RF-based accelerators, where the photo-emitting laser pulse is synchronized to the RF-drive by standard RF-techniques (\textit{i.e.} phase locked loops operating at GHz speeds).} This proof-of-principle THz linear accelerator demonstrates the potential for an all-optical acceleration scheme that can be readily integrated into small-scale laboratories providing users with electron beams that will enable new experiments in ultra-fast electron diffraction and x-ray production.

\begin{methods}
%Put methods in here.  If you are going to subsection it, use
%\verb|\subsection| commands.  Methods section should be less than
%800 words and if it is less than 200 words, it can be incorporated
%into the main text.

\subsection{THz Pulse Generation} The THz pulse which is used in the accelerating structure is generated with optical rectification of 700~fs, 1.03~$\mu$m pulses in cryogenically cooled stoichiometric Lithium Niobate (LN) doped with 1$\%$ MgO. LN was chosen because it exhibits multiple advantages with low THz absorption, large bandgap, a high damage threshold and a high effective nonlinear coefficient, \textit{i.e.} large d$_{\textrm{eff}}$. For a high-efficiency of conversion in LN a tilted-pulse-front pumping scheme\cite{huang2013high} is required. The THz pulse is centered at 0.45~THz with a broad spectrum ranging from  0.2 to 0.8~THz. A THz pulse energy of 10~$\mu$J is produced from 1.2~mJ of NIR which is slightly lower than the peak conversion efficiency of THz generation\cite{huang2013high} due to a larger spot size in the LN (decreased fluence) for improved transport of the THz beam. The THz beam has excellent Gaussian mode content which allows for low-loss coupling. The THz beam was characterized spatially by a pyroelectric detector array (Spiricon Pyrocam IV, Ophir Photonics). 

\subsection{Pump source}
The pump source for THz generation is a Yb:KYW chirped pulse regenerative amplifier (RGA) producing 1.5~mJ pulses with 1 kHz repetition rate at a center wavelength of 1030~nm and bandwidth of 2.1~nm. The dielectric grating compressor following the RGA compresses the pulses to a near transform-limited 700~fs pulse duration (FWHM). The seed for the RGA was a mode-locked Yb-doped fiber oscillator emitting 70~fs, 0.2~nJ pulses at 80~MHz \cite{hong2008generation} amplified to 1.6~nJ by a Yb-doped fiber amplifier. After losses through the optical elements in the pulse front tilting setup including a diffraction grating, the impinging pump energy into the (LN) crystal was 1.2 mJ.

\subsection{UV Photoemitter}
About 2$\%$ of the available 1030~nm NIR pump energy was used to generate the UV photoemitter pulses by fourth-harmonic generation from 1.03~$\mu$m based on two-stage second harmonic generation (SHG) in a 10-mm-long type I LBO crystal for IR to green and in a 0.5-mm-long type-I BBO crystal for the green to UV, respectively. A BG-39 bandpass filter was used to remove the fundamental NIR component after the first stage. The UV pulses have a duration of $\sim$350~fs and are focused onto the photocathode with a beam waist of $\sim$200~$\mu$m. Since both the THz and UV pulses are produced from the same sub-ps NIR laser, the timing jitter between them is negligible.

\subsection{Electro-optic Sampling} Electro-optic (EO) sampling was used to determine the temporal and spectral properties of the THz pulse and dispersion induced from the quasi-optical elements in the THz beamline. Optical synchronization between the THz pulse and the mode-locked Yb-doped fiber seed oscillator (80~MHz, 70~fs, 1030~nm) was ensured as explained in the ``Pump source" section of ``Methods". Birefringence was induced in a 200~$\mu$m thick, 110-cut ZnTe crystal. A quarter-wave plate followed by a polarizer converts the field-induced birefringence to an intensity modulation, and the intensity modulation is measured using a balanced detector with a delay scan. 

\subsection{Radially Polarized THz Beam} The THz pulse generated by optical rectification is linearly polarized, which is not compatible with the TM$_{01}$ mode used in the accelerating structure. A segmented half waveplate was used to convert the linearly-polarized light to radially-polarized light\cite{grosjean2008linear}. Each segment of the waveplate imparts the appropriate rotation to the polarization to transition from a linear to a radially-polarized beam which couples well in the far field to the TM$_{01}$ mode of the accelerating structure. A segmented $\lambda/2$ waveplate with 8 segments of $\sim$8~mm thick quartz designed for operation at 0.45~THz was used.

\subsection{Structure Testing} Transmitted energy and polarization measurements were performed to test and optimize the coupler and waveguide performance. Measurements were performed with a Gentec-EO Pyroelectric Joulemeter Probe that is capable of measuring pulse energies exceeding 100~nJ. Efficient excitation of the TM$_{01}$ mode, demonstrated in these measurements, is a critical requirement in developing a compact high-gradient THz LINAC. The vertical and horizontal polarization measured after the segmented waveplate were 53$\%$ and 47$\%$, respectively. The waveguide length was 5~cm, including two tapers, demonstrating that ohmic losses are manageable even over significant interaction lengths. With a measured energy of 2~$\mu$J at the exit of the waveguide the calculated on-axis electric field is 9.7~MV/m.

\subsection{{THz LINAC}}

{The THz pulse accelerates electrons in a circular waveguide consisting of a quartz capillary inserted into a hollow copper cylinder, see Figure 1(a). The inner diameter of the copper waveguide is 940~$\mu$m with a dielectric wall thickness of 270~$\mu$m. This results in a vacuum space with a diameter of 400~$\mu$m. The dielectric constant of the quartz capillary is nominally $\varepsilon = 4.41$. The operational mode of the LINAC is a traveling TM$_{01}$ mode, see Figure 4(a). The dispersion relation for the operating mode is shown in Figure 4(b). At the center frequency of the THz pulse (450~GHz) the group velocity is $v_g/c=0.46$ and the phase velocity is $v_p/c=0.505$. Due to the operational frequencies proximity to the cutoff of the waveguide, the mode is highly dispersive with the phase and group velocity shown in Figure 4(c) as a function of frequency. The waveguide dimensions of the LINAC were chosen to optimize for an experimental setup with a low initial electron energy of 60~keV, the THz pulse energy available and the transverse dimension of the electron beam. At the nominal 60~keV the electron velocity is $v/c=0.45$ and the mismatch with the phase velocity results in an interaction length (i.e. slippage equal to the THz wavelength) of 3~mm. }

{With increased THz energy and increased electron energy one can consider a relativistic accelerator design in which the phase velocity of the TM$_{01}$ mode is equal to the speed of light. Additionally, the design of the waveguide can be optimized to match the frequency of the available source. Figures 5(a-e) present the frequency of operation, energy gain, acceleration gradient, group velocity and interaction length as a function of vacuum radius and dielectric thickness assuming a 10~mJ single-cycle THz pulse and an initial electron energy of 1~MeV. Figure 5(f) presents the electron energy as a function of distance for two cases which operate with a frequency of 0.45/1~THz, a vacuum space with a diameter of 210/210~$\mu$m and a dielectric wall thickness of 90/30~$\mu$m. Further numerical studies for relativistic THz LINACs using dielectric-loaded waveguides and single cycle THz pulses were previously reported by Wong et al.\cite{wong2013compact} }

\end{methods}

\bibliography{PhDbib}

%% Here is the endmatter stuff: Supplementary Info, etc.
%% Use \item's to separate, default label is "Acknowledgements"

\begin{addendum}
 \item The authors would like to acknowledge W. S. Graves for many discussions and help with Parmela simulations; L. J. Wong for discussions on THz waveguides used for acceleration; S.-W. Huang for his assistance with THz generation; and L. Hua for his assistance with the operation of the laser system. This work was supported by DARPA under contract N66001-11-1-4192, by AFOSR under grant AFOSR - A9550-12-1-0499, DOE DE-FG02-10ER46745, DOE DE-FG02-08ER41532, ERC Synergy Grant 609920 and NSF DMR-1042342, the Center for Free-Electron Laser Science at DESY and the excellence cluster “The Hamburg Centre for Ultrafast Imaging- Structure, Dynamics and Control of Matter at the Atomic Scale” of the Deutsche Forschungsgemeinschaft.

 \item[Competing Interests] The authors declare that they have no competing financial interests.
 \item[Correspondence] Correspondence and requests for materials should be addressed to F. X. K{\"a}rtner~(email: franz.kaertner@cfel.de).
 \item[Author Contributions] E.A.N., A.F. and F.X.K. conceived and designed the experiment.
E.A.N. built and performed the experiment with help from W.R.H. and K-H.H. 
A.F. and E.A.N. provided the design for the dielectric-loaded metal waveguide.
W.R.H. and K.R. provided the terahertz source.
G.M. and R.J.D.M. were responsible for the DC gun.
W.R.H., K.R., K-H.H., A.F., and F.X.K. provided feedback to improve the experiment.
E.A.N. developed and performed the simulations for data evaluation and interpretation of results.
E.A.N. wrote the manuscript with contributions from W.R.H., K.R. and K-H.H and revisions by all.
K-H.H. and F.X.K. provided management and oversight to the project.

\begin{figure}
  \centering
	  \includegraphics[width=1\textwidth]{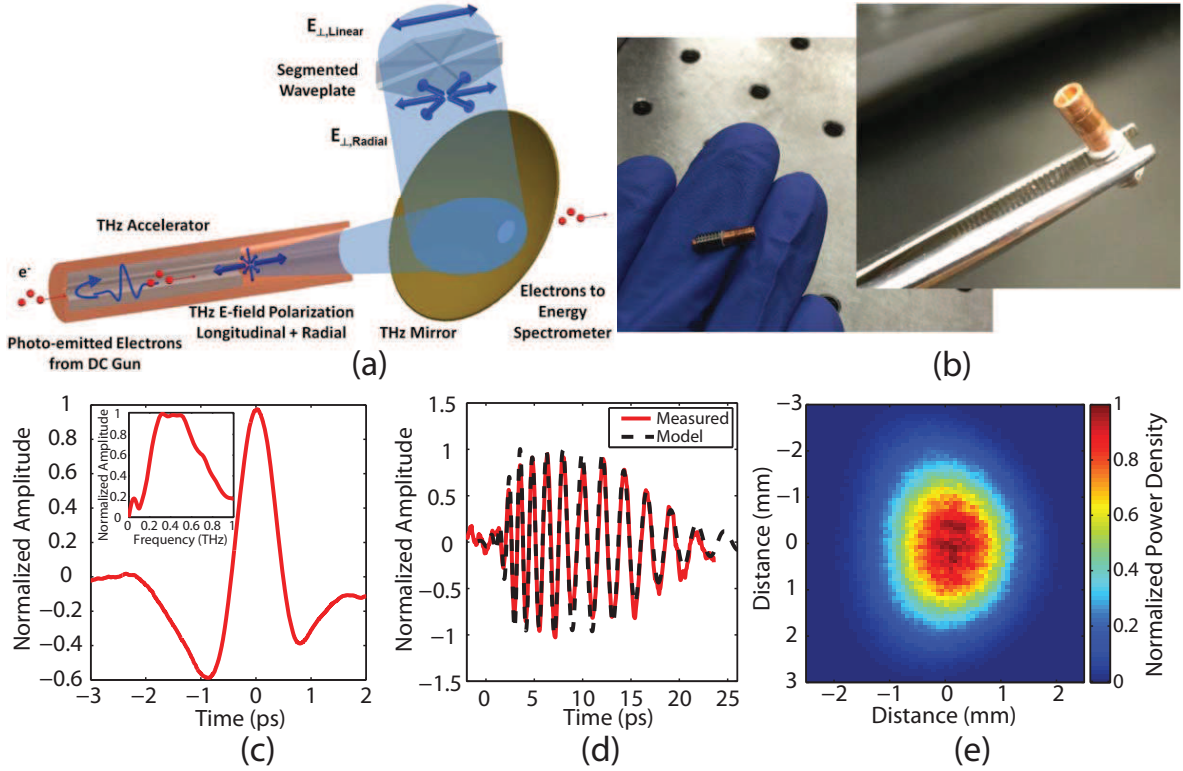}

	%\caption {(a) Normalized intensity of the focused THz beam and (b) the time-domain waveform of the THz pulse determined with EO sampling. Insert: Corresponding frequency-domain spectrum.}
  %\label{fig:THzpulse}
	\caption {\textbf{Terahertz-driven linear accelerator} {(a) Schematic of the THz LINAC. (Top Right) A linearly polarized THz pulse is converted into a radially polarized pulse by a segmented waveplate before being focused into the THz waveguide. The THz pulse is reflected at the end of the waveguide to co-propagate with the electron bunch which enters the waveguide through a pin-hole (Lower Left). The electron bunch is accelerated by the longitudinal electric field of the co-propagating THz pulse. The electron bunch exits the THz waveguide and passes through a hole in the focusing mirror (Right) for the THz pulse. }(b) Photograph of the compact mm-scale THz LINAC. {(c) The time-domain waveform of the THz pulse determined with EO sampling (See Methods: Electro-optic Sampling). Insert: Corresponding frequency-domain spectrum.} (d) The time-domain waveform of the THz pulse at the exit of a 4~cm THz waveguide. (e) Normalized intensity of the focused THz beam. }
  \label{fig:THzpulse}
	%\vspace{-7mm}
\end{figure}

\begin{figure}
  \centering
		\includegraphics[width=0.6\textwidth]{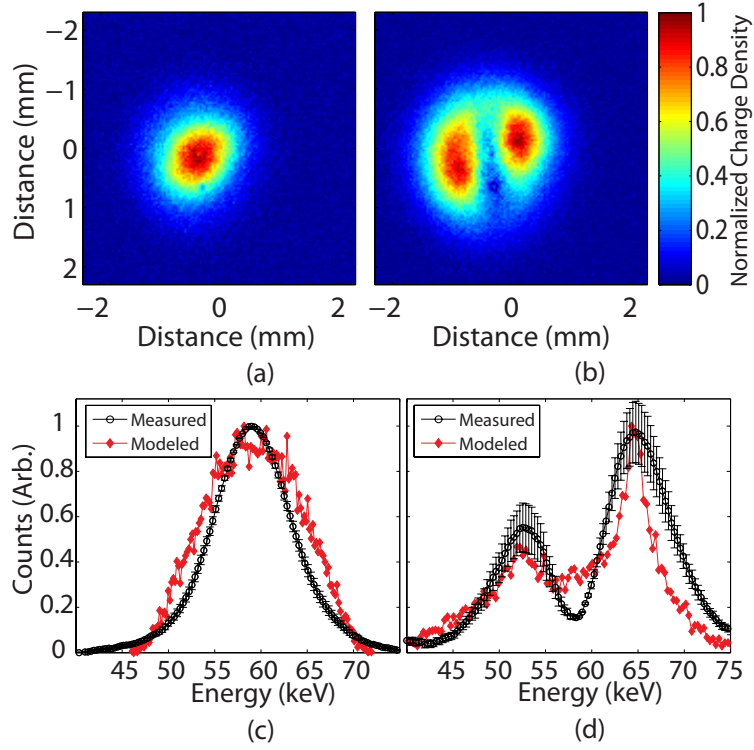}
  
		\caption {\textbf{Demonstration of terahertz acceleration}  {Transverse electron density of the electron bunch as recorded by an MCP at 59~kV for (a) THz off and (b) THz on.} The bimodal distribution is due to the presence of accelerated and decelerated electrons which are separated spatially by the magnetic-dipole energy spectrometer. The images are recorded over a 3~s exposure at 1~kHz repetition rate. {(c) Comparison between simulated (red) and measured (black) energy spectrum of the electron bunch measured at the MCP due to the deflection of the beam by a magnetic dipole. At 59~keV and with 25~fC per bunch, the simulation predicts a $\sigma_\perp=513~\mu$m and $\Delta E/E=1.25$~keV. The observed $\Delta E/E$ appears larger due to the large transverse size of the beam.  After the pin-hole the transverse emittance is 25~nm-rad and the longitudinal emittance is 5.5~nm-rad.} (d) Comparison between simulated (red) and measured (black) electron bunch at MCP after acceleration with THz. Decelerated electrons are present due to the long length of the UV pulse which generates the electron bunch. }
  \label{fig:energygain}
\end{figure}

\begin{figure}
	\includegraphics[width=1\textwidth]{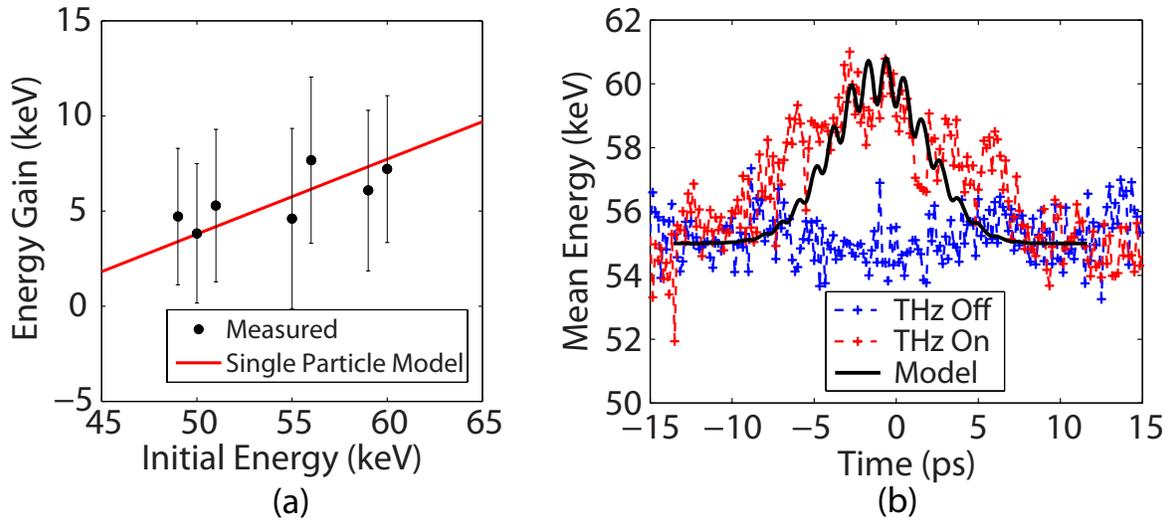}
		\caption {\textbf{Acceleration gradient and terahertz phasing} {(a) Scaling of energy gain for accelerated electrons as a function of the initial electron energy at the entrance of the THz LINAC. Black dots with one sigma error bars are measured values  and the red line is a single particle model.} {(b) The temporal profile for the mean energy gain of accelerated electrons comparing the THz on and THz off signal against the simulated electron bunch. The initial electron energy was set at 55~keV to ensure stable performance of the DC electron gun over the acquisition time of the data set.}}
  \label{fig:scans}
\end{figure}

\begin{figure}
  \centering
		\subfloat[]{\includegraphics[width=0.24\textwidth]{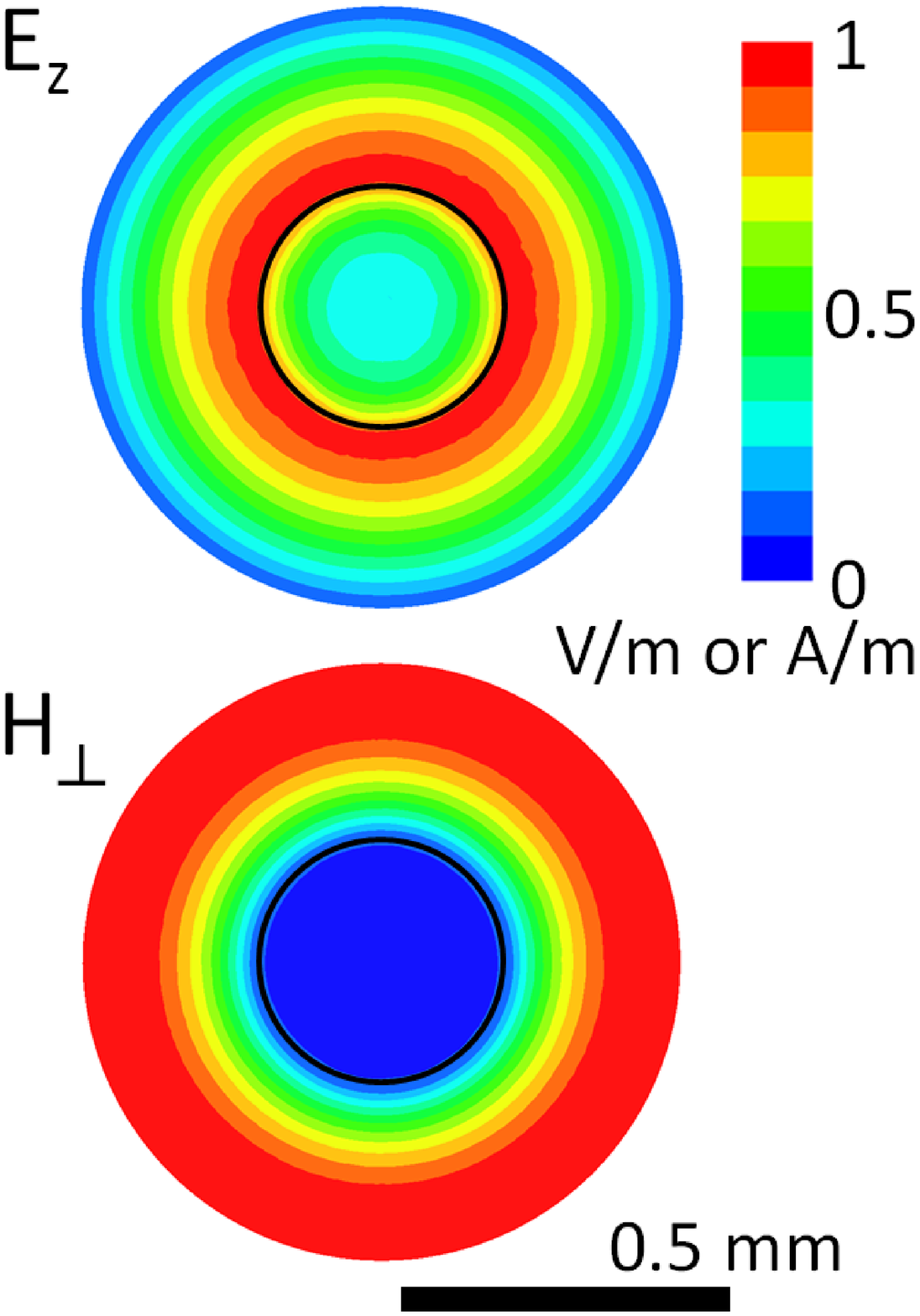}
		%7/6/2013
	\label{fig:mode}}
  \subfloat[]{\includegraphics[width=0.36\textwidth]{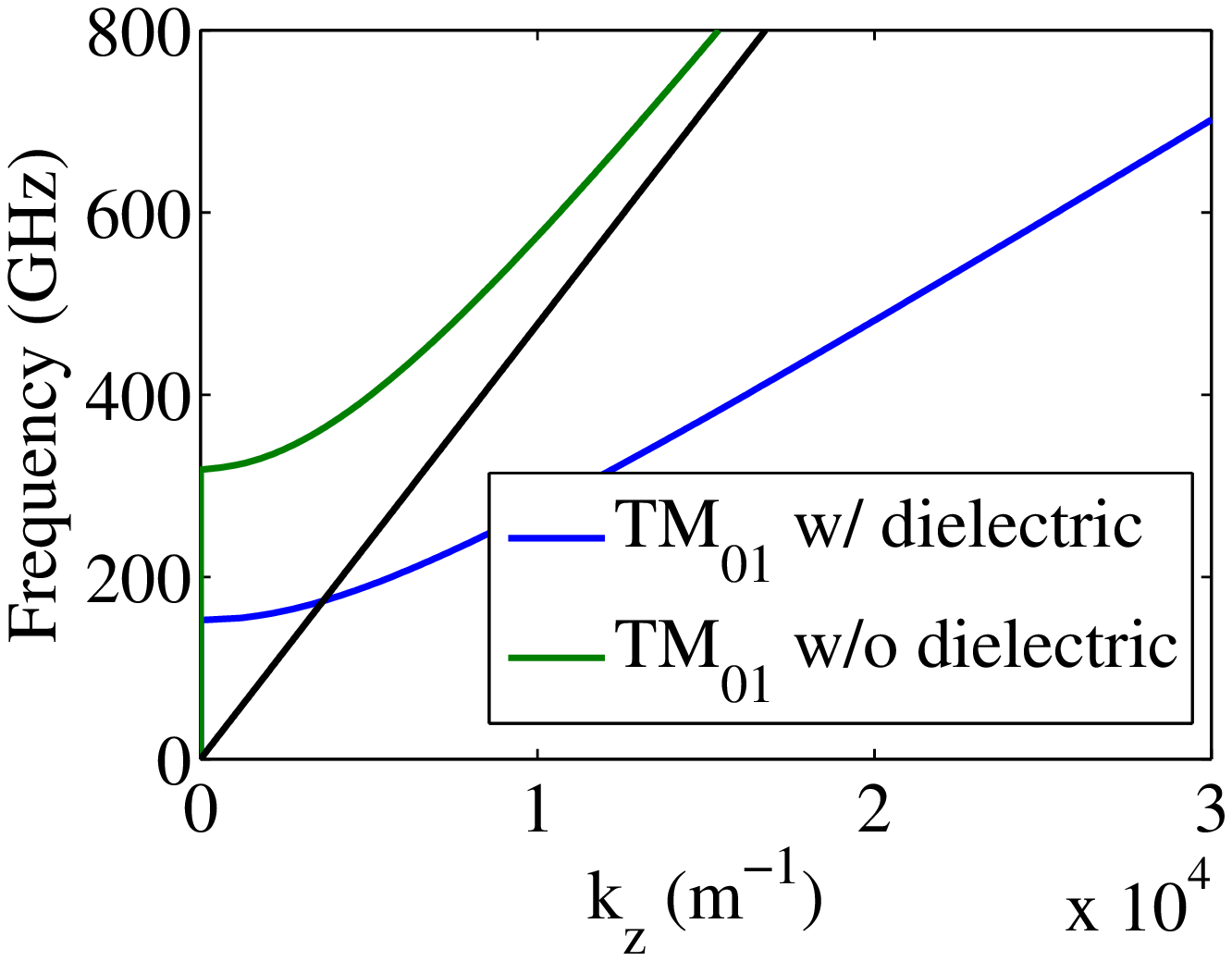}
  \label{fig:disp}}
	\subfloat[]{\includegraphics[width=0.36\textwidth]{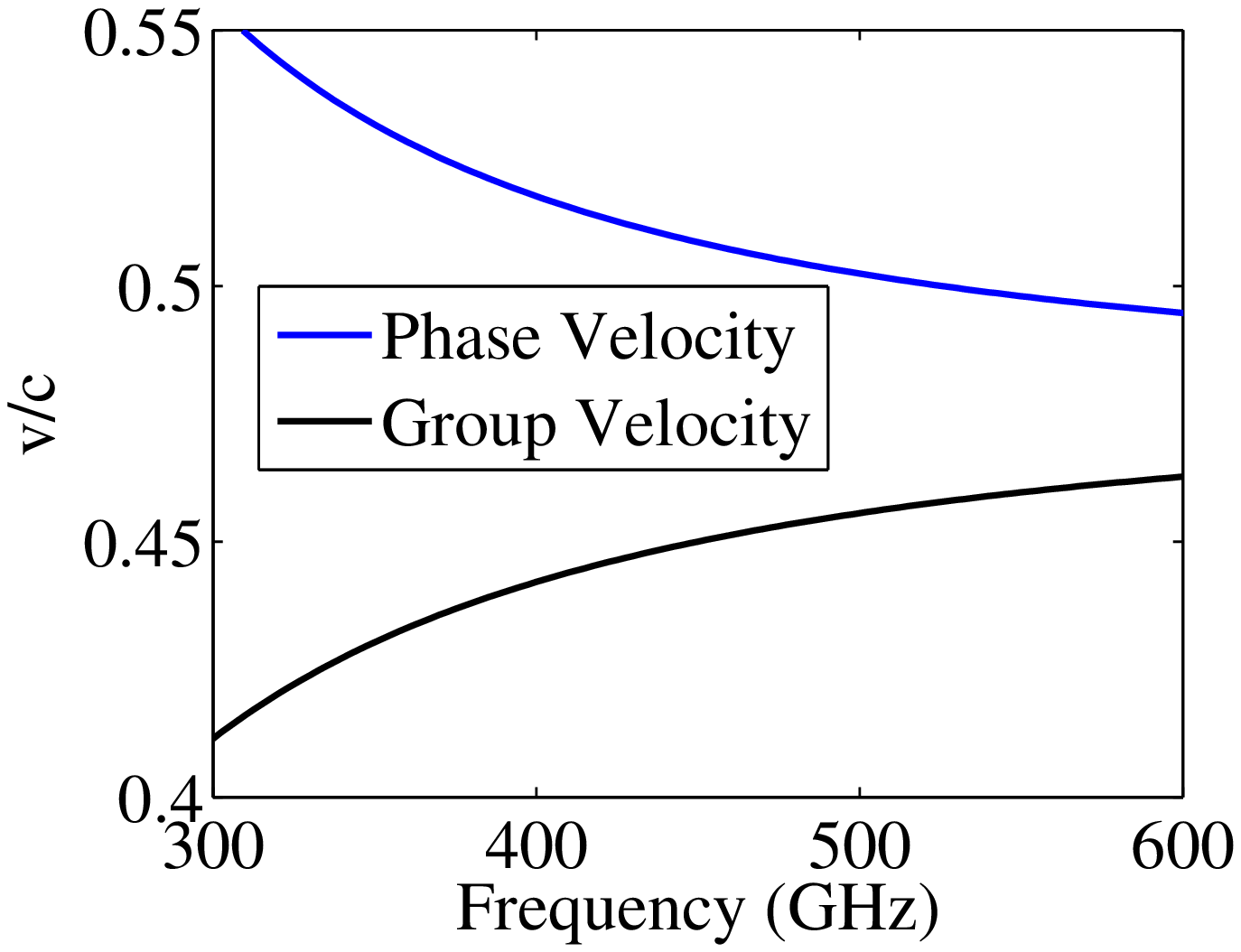}
	\label{fig:vel}}
	\caption {{\textbf{TM$_{01}$ THz LINAC Parameters} (a) Normalized magnitude of the longitudinal electric field and perpendicular magnetic field for the TM$_{01}$ mode at 450 GHz in a circular copper waveguide with dielectric loading. The inner diameter of the copper waveguide is 940~$\mu$m with a dielectric wall thickness of 270~$\mu$m. The solid black line indicates the boundary between the vacuum core and the quartz capillary.  (b) The dispersion relation for the TM$_{01}$ mode with and without dielectric loading. The black line indicates the speed of light in vacuum. (c) The group and phase velocity of the THz pulse as a function of frequency with dielectric loading.}}
  \label{fig:nonrelativistic}
\end{figure}

\begin{figure}[!ht]
  \centering
	\subfloat[]{\includegraphics[width=0.32\textwidth]{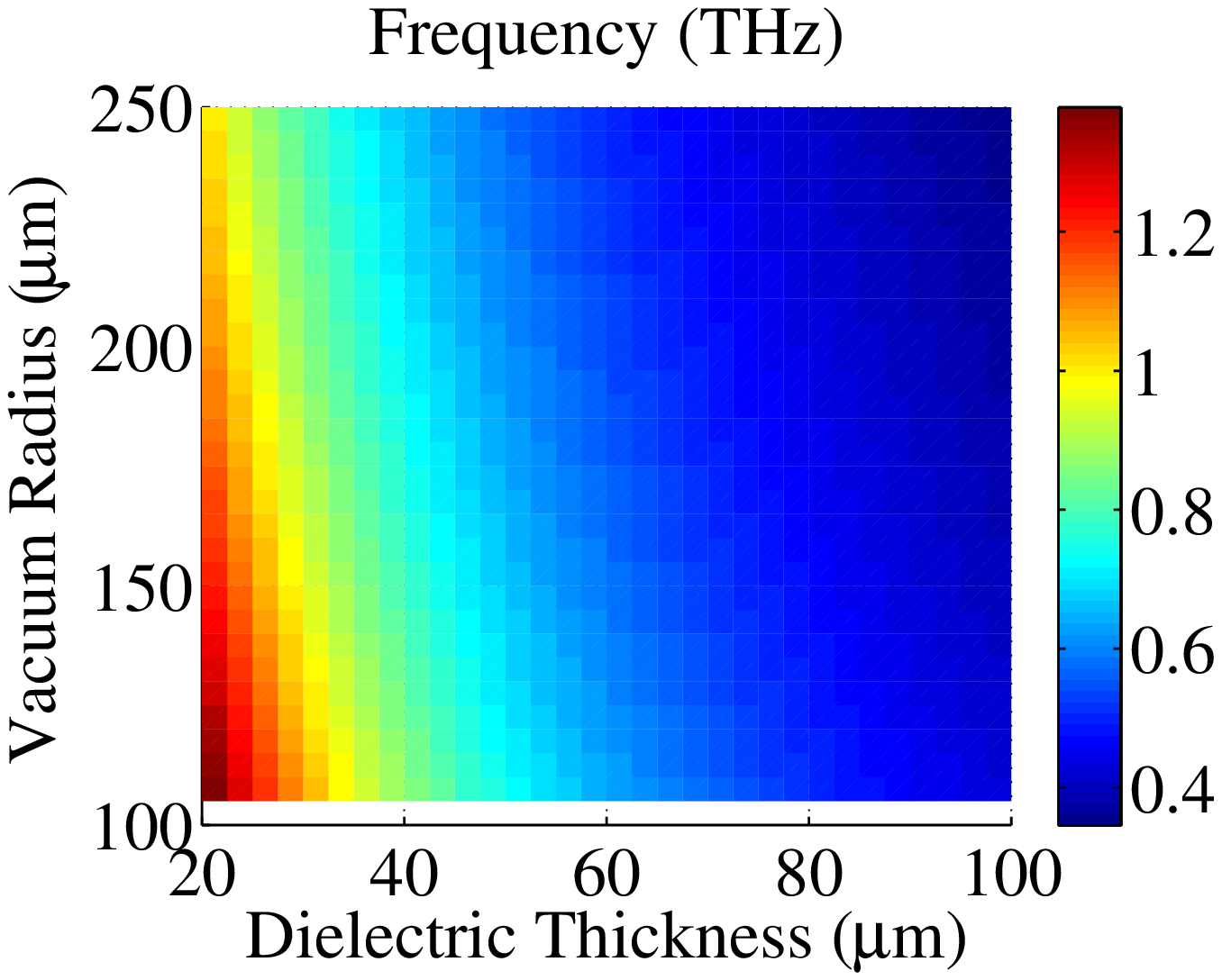}
\label{fig:freq}}
	%\hfil
 \subfloat[]{\includegraphics[width=0.32\textwidth]{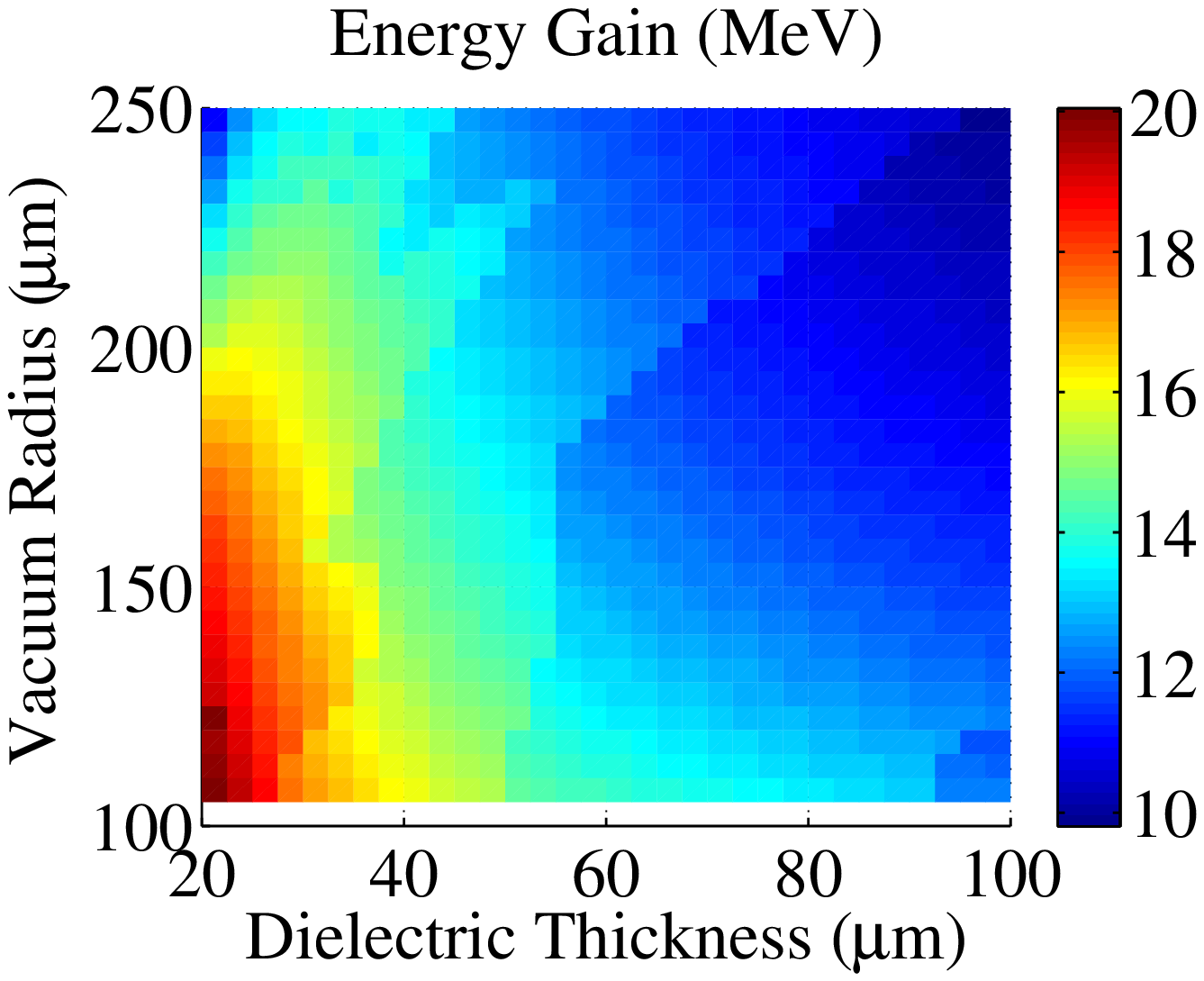}
  \label{fig:EG}}
	%\hfil
	\subfloat[]{\includegraphics[width=0.32\textwidth]{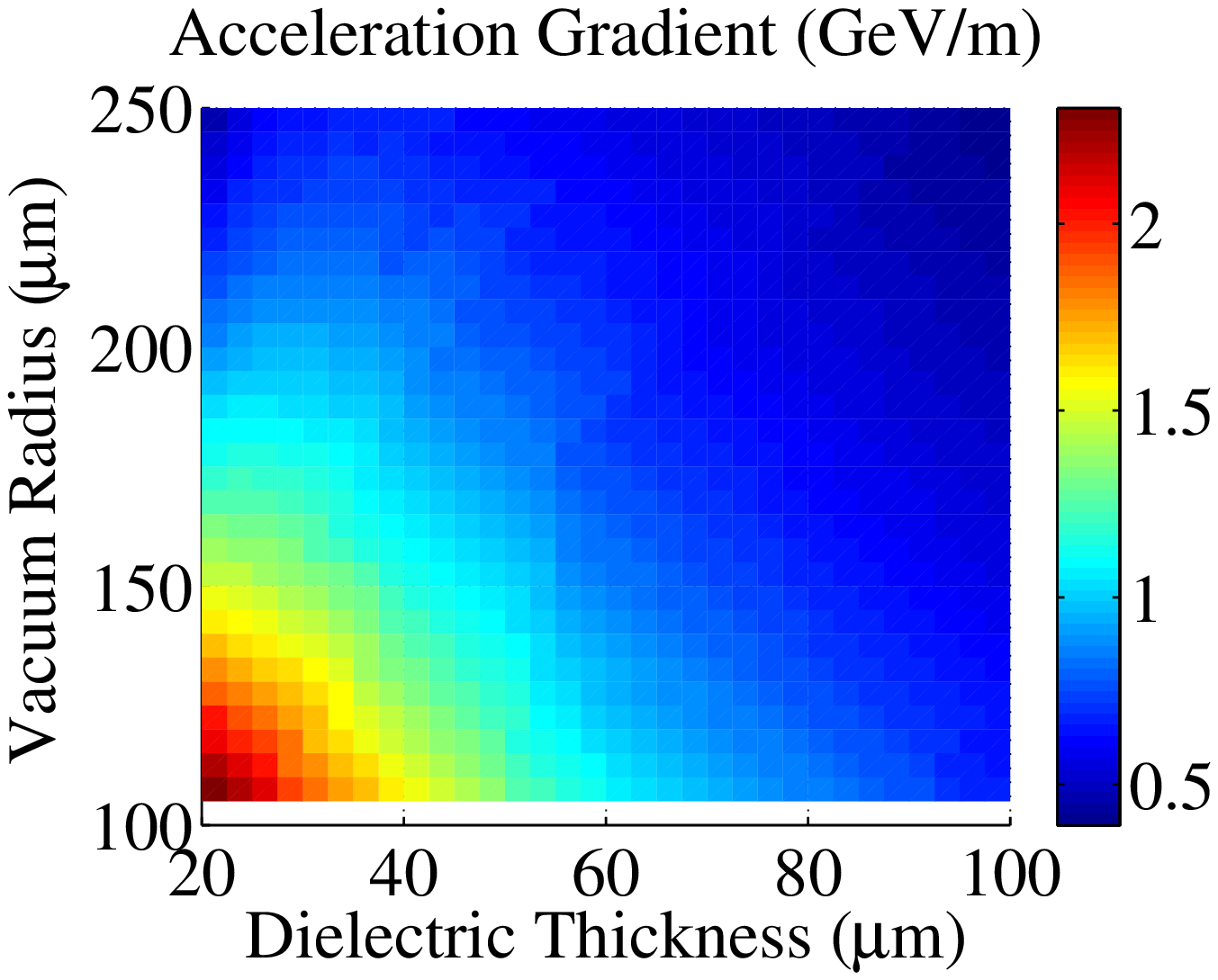}
	\label{fig:AG}}\\
\subfloat[]{\includegraphics[width=0.32\textwidth]{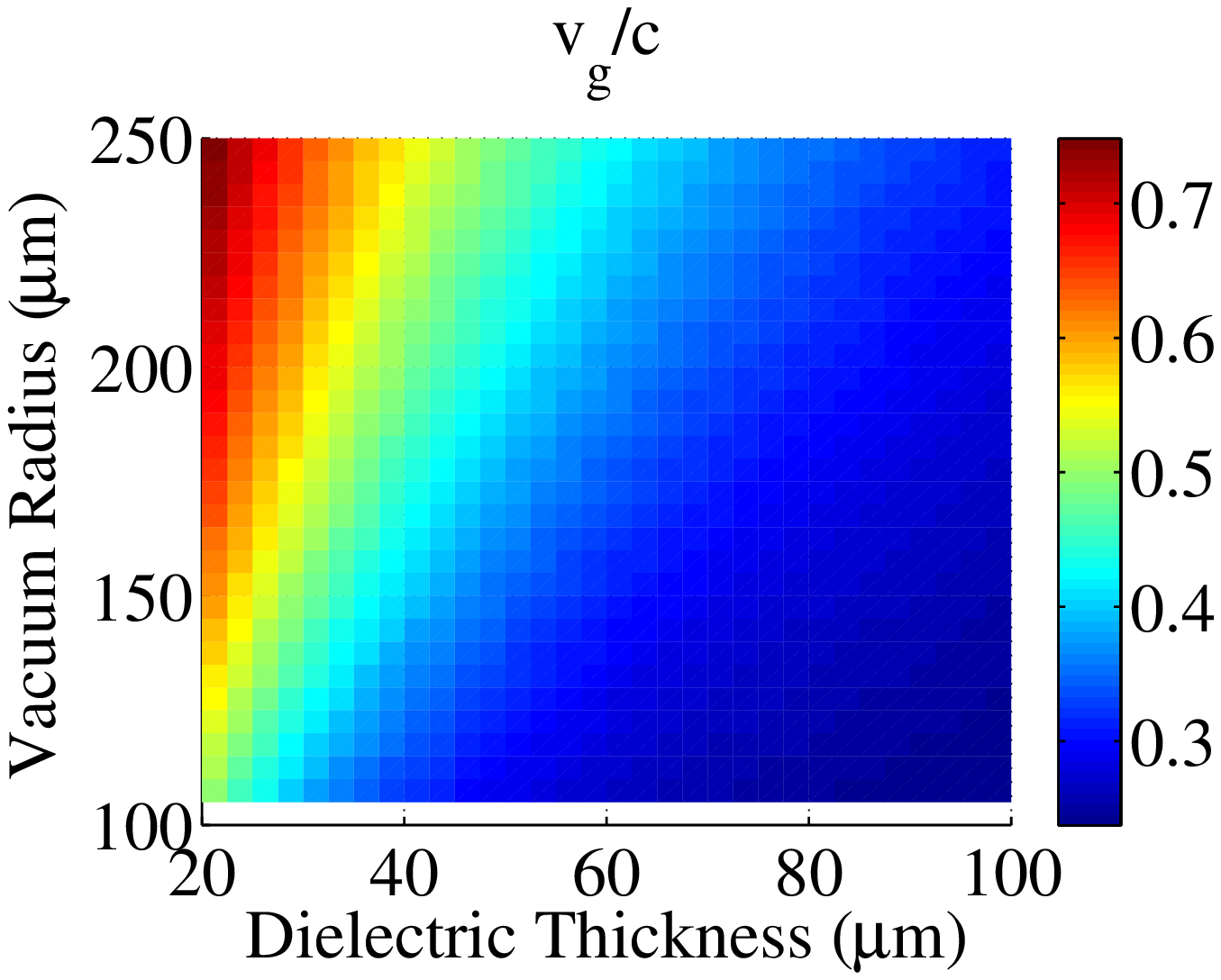}
	\label{fig:GV}}
	%\hfil
 \subfloat[]{\includegraphics[width=0.32\textwidth]{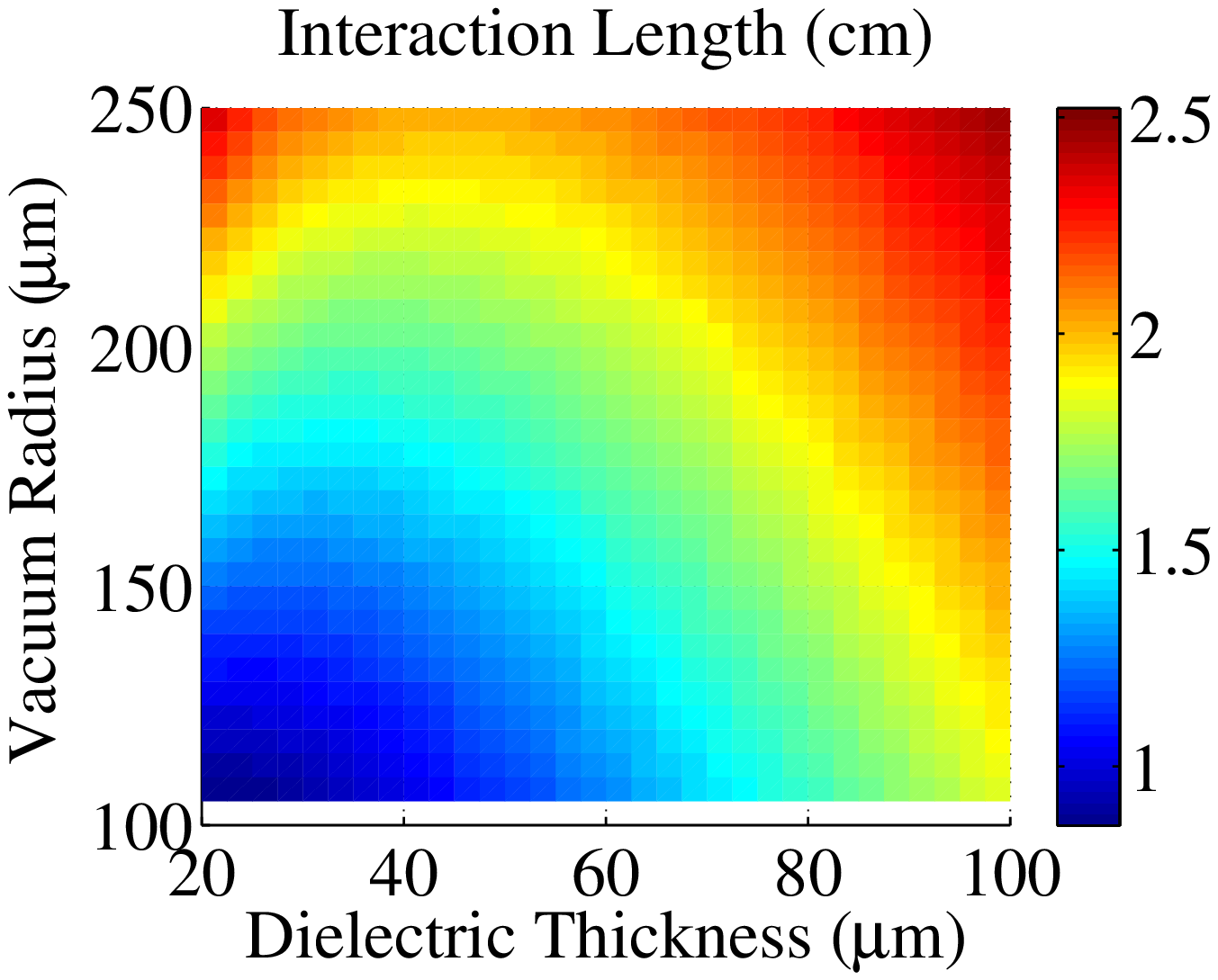}
  \label{fig:IL}}
	%\hfil
	\subfloat[]{\includegraphics[width=0.32\textwidth]{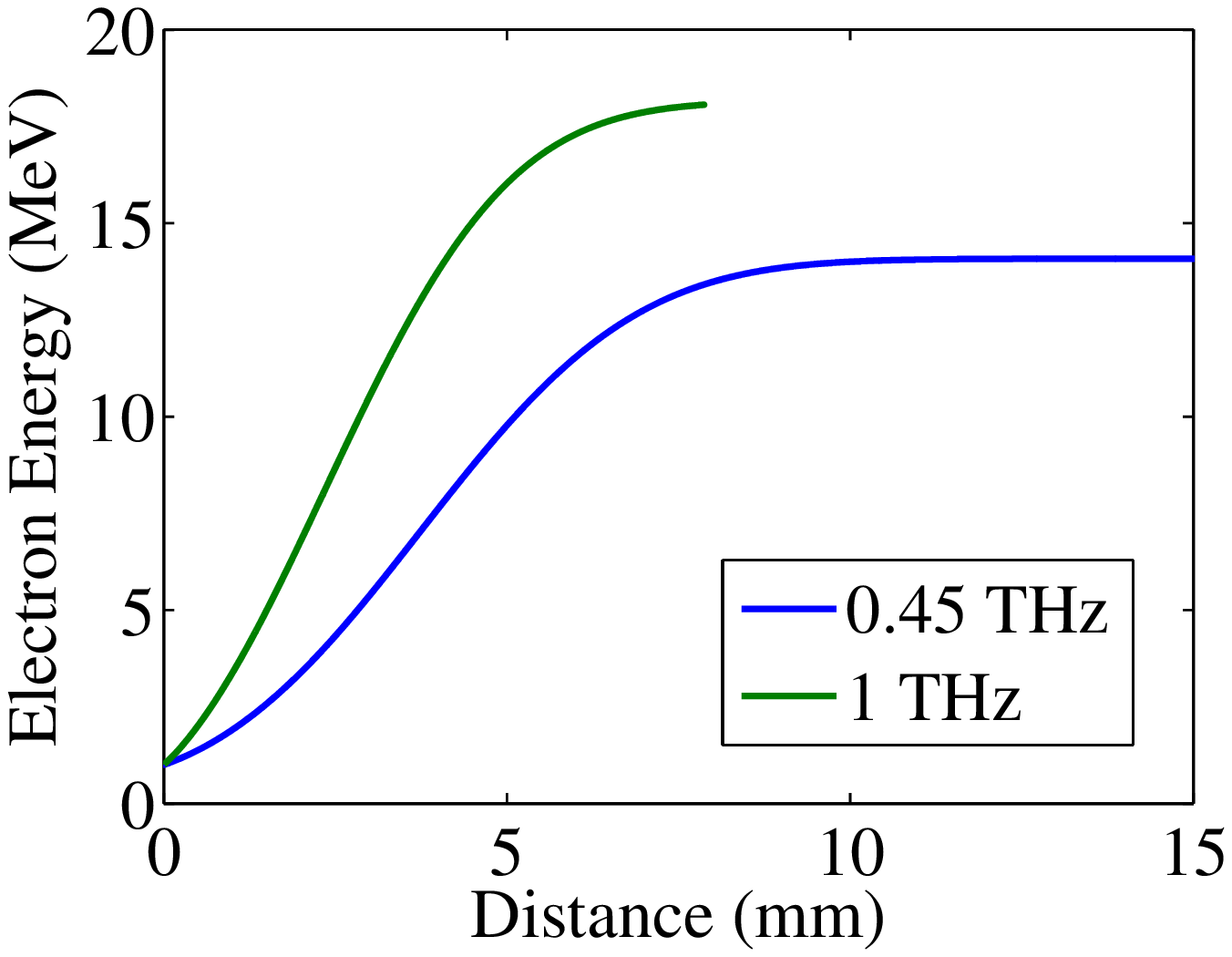}
	  \label{fig:ED}}
	\caption {{\textbf{Relativistic THz LINAC Design} Performance parameters as a function of vacuum radius and dielectric wall thickness for a relativistic THz LINAC operating in the TM$_{01}$ mode with a 10~mJ single-cycle THz pulse and an initial electron energy of 1~MeV. The phase velocity is $c$ for the nominal frequency of operation. The (a) frequency of operation, (b) energy gain, (c) acceleration gradient, (d) group velocity and (e) interaction length for the THz LINAC. (f) The electron energy as a function of distance for two cases which operate with a frequency of 0.45/1~THz, a vacuum space with a diameter of 210/210~$\mu$m and a dielectric wall thickness of 90/30~$\mu$m. }}
  \label{fig:relativistic}
\end{figure}

\end{addendum}

%%
%% TABLES
%%
%% If there are any tables, put them here.
%%

\end{document}